\documentclass[10pt]{iopart}
\usepackage{iopams}
\usepackage{cite}
\usepackage{bm}
\usepackage{amssymb}

\newtheorem{proposition}{Proposition}
\newtheorem{definition}{Definition}

\begin{document}

\title[Direction of light propagation to order $G^2$]{Direction of light propagation to order $G^2$ in static,\\
spherically symmetric spacetimes: a new derivation}

\author{Pierre Teyssandier}

\address{SYRTE, CNRS/UMR 8630, UPMC, Observatoire de Paris, 61 avenue de \\
 l'Observatoire, F-75014 Paris, France}

\ead{Pierre.Teyssandier@obspm.fr}

\begin{abstract}

A procedure avoiding any integration of the null geodesic equations is used to derive the direction of light propagation in a three-parameter family of static, spherically symmetric spacetimes within the post-post-Minkowskian approximation. Quasi-Cartesian isotropic coordinates adapted to the symmetries of spacetime are systematically used. It is found that the expression of the angle formed by two light rays as measured by a static observer staying at a given point is remarkably simple in these coordinates. The attention is mainly focused on the null geodesic paths that we call the quasi-Minkowskian light rays. The vector-like functions characterizing the direction of propagation of such light rays at their points of emission and reception are firstly obtained in the generic case where these points are both located at finite distances from the centre of symmetry. The direction of propagation of the quasi-Minkowskian light rays emitted at infinity is then straightforwardly deduced. An intrinsic definition of the gravitational deflection angle relative to a static observer located at a finite distance is proposed for these rays. The expression inferred from this definition extends the formula currently used in VLBI astrometry up to the second order in the gravitational constant $G$. 

\end{abstract}


\maketitle

\section{Introduction} \label{Sintro}

Testing gravity with projects like LATOR \cite{Turyshev:2009}, ASTROD \cite{Braxmaier:2011} and others will require a fully relativistic modelling of the direction of light propagation up to\footnote{In the present work, ``up to" always means ``up to and including".} terms of order $G^2$, with $G$ being the Newtonian gravitational constant (see, e.g., \cite{Ashby:2010,Minazzoli:2011} and references therein). Such modelling appears to be relevant even for the Gaia mission \cite{Turon:2005}, yet usually treated within the post-Minkowskian or post-Newtonian approximations (\cite{Kopeikin:1999,Kopeikin:2002,Klioner:2003,Crosta:2011} and references therein). Indeed, an analytical solution for the direction of light propagation in the post-linear regime has been recently derived for a three-parameter family of static, spherically symmetric spacetimes with the view of explaining an apparent discrepancy between the standard post-Newtonian approach of Gaia and the numerical integration of the differential equations of the null geodesics \cite{Klioner:2010}. Although being perfectly efficient, this solution presents the disadvantage of inferring the direction of propagation of light at its emission and reception points from the solution which corresponds to a ray emitted at infinity in a given direction. Our aim, in the present paper, is to furnish a new, more direct derivation based on a procedure developed in \cite{Leponcin:2004} and \cite{Teyssandier:2008}. This derivation is natively adapted to the generic case where both the emitter and the receiver of the light ray are located at a finite distance from the source of the field. Furthermore, no integration of the null geodesic equations is needed. 

The paper is organized as follows. Section \ref{Snot} lists the notations and conventions we use. Sections \ref{Spropa} and \ref{Sangle} yield the general formulae which characterize the direction of light propagation at the points of emission and reception in any static, spherically symmetric spacetime. The metric is assumed to be asymptotically flat and written for convenience in isotropic, quasi-Cartesian coordinates adapted to the symmetries of spacetime. In section \ref{Spropa} it is recalled how the direction of propagation and the impact parameter of a light ray can be derived from the time transfer function giving the travel time of a photon as a function of the spatial coordinates of the emitter and the receiver. In section \ref{Sangle} it is shown that the angle between two light rays passing through a given point as measured by an observer at rest at this point is equal to the Euclidean angle formed by the triples characterizing the directions of the rays when isotropic coordinates are used. In the subsequent sections it is assumed that the components of the metric can be expanded in power series of $m/r$, where $m=GM/c^2$, with $M$ being the mass of the central body and $r$ the isotropic radial variable. The expansions are systematically stopped at order $m^2/r^2$. Furthermore, only the null geodesic paths for which the parametric equations can be expanded in power series of $G$ are considered. These paths are called {\it quasi-Minkowskian light rays} in the present work. Section \ref{SppNap} is devoted to the calculation of the impact parameter and the direction of light propagation up to order $G^2$ in the generic case where both the emission and reception points are located at a finite distance from the centre of the gravitational source. The case of a ray emitted at infinity is treated in section \ref{Sinfty}. Section \ref{Sbend} yields an intrinsic expression of the gravitational bending of a light ray coming from infinity and observed by a static observer at a given point. This expression is the post-post-Minkowskian extension of the well-known formula currently used in VLBI astrometry. The classical expression of the light deflection in the post-linear regime is recovered in section \ref{Stdef}. The conclusion is given in section \ref{Conclusion}. An appendix is devoted to the conservation law of the angular momentum in isotropic, quasi-Cartesian coordinates.

\section{Notations and conventions} \label{Snot}

 We use the following notations and conventions.

\begin{itemize}
\item
Spacetime is assumed to be covered by a global quasi-Cartesian coordinate system $(x^{\mu})=(x^0,{\bm x})$, where $x^0=ct$, $t$ 
being a time coordinate, and ${\bm x}=(x^i)$.

\item
Greek indices run from 0 to 3, and latin indices run from 1 to 3.

\item 
The signature adopted for the metric is $+ - - -$.

\item
Any bold italic letter refers to an ordered triple. We distinguish the triples built 
with contravariant components of a 4-vector from the ones built with covariant components by 
systematically using the notations $(a^1,a^2,a^3)=(a^i)=\bm a$ and $(b_1,b_2,b_3)=(b_i)=\underline{\bm b}$. All these triples are regarded as 3-vectors of the ordinary Euclidean space.

\item
Given three triples $\bm a$, $\underline{\bm b}$ and $\underline{\bm c}$, $\underline{\bm c}=\bm a+\underline{\bm b}$ means that $c_i=a^i+b_i$ for any $i=1,2,3$. Such a mixing 
of `contravariant' and `covariant' quantities cannot be ambiguous, since these equalities must be understood as 
equalities of triples, and not as equalities between true 4-vectors. 

\item
Given three triples $\bm a, \bm b$ and $\underline{\bm c}$, $\bm a . \bm b$ denotes the Euclidean scalar product $a^i b^i$ and $\bm a . \underline{\bm c}$ denotes $a^i c_i$, Einstein's convention on repeated indices being used in both cases. 

\item
$\vert \bm a \vert$ denotes the formal Euclidean norm of the triple $\bm a$: $\vert \bm a \vert=(\bm a . \bm a)^{1/2}$. Similarly,  $\vert \underline{\bm b}\vert = (\underline{\bm b} . \underline{\bm b})^{1/2}$. If $\vert \bm a \vert=1$, $\bm a$ is conventionally called a unit (Euclidean) 3-vector.

\item
$\bm a\times\bm b$ and $\bm a\times\underline{\bm c} $ are the triples obtained by the usual rule giving the exterior product of two vectors of the Euclidean space.

\item
The Euclidean angle between two triples $\bm a$ and $\bm b$ is denoted by $\delta_{\scriptscriptstyle {\cal E}}(\bm a, \bm b)$. By convention, the determination of this angle is chosen so as to have $0\leq\delta_{\scriptscriptstyle {\cal E}}(\bm a, \bm b)\leq\pi$. The angle $\delta_{\scriptscriptstyle {\cal E}}(\bm a, \bm b)$ can be calculated by the relation $\delta_{\scriptscriptstyle {\cal E}}(\bm a, \bm b)=\arccos(\bm a . \bm b/\vert\bm a\vert . \vert\bm b\vert)$. The same notation and convention hold when $\bm a$ or $\bm b$ are replaced by `covariant' triples $\underline{\bm c}$ or $\underline{\bm d}$. Thus $\delta_{\scriptscriptstyle {\cal E}}(\bm a, \underline{\bm c})=\arccos(\bm a . \underline{\bm c}/\vert\bm a\vert . \vert\underline{\bm c}\vert)$ and $\delta_{\scriptscriptstyle {\cal E}}(\underline{\bm c}, \underline{\bm d})=\arccos(\underline{\bm c} . \underline{\bm d}/\vert\underline{\bm c}\vert . \vert\underline{\bm d}\vert)$.

\item
Given two 4-vectors $V$ and $W$, $V\cdot W$ stands for the Lorentzian scalar product of $V$ and $W$: $V\cdot W=g(V,W)=g_{\mu\nu}V^{\mu}W^{\nu}$, the quantities $g_{\mu\nu}$ being the components of the metric tensor $g$.
\end{itemize}

\section{Direction of light propagation in a static, spherically symmetric spacetime} \label{Spropa}

Throughout this paper, spacetime is assumed to be a static, spherically symmetric Lorentzian manifold. For convenience, the metric is written in isotropic coordinates $(x^0, \bm x)$ adapted to the symmetries of spacetime:
\begin{equation} \label{ds2}
ds^2={\cal A}(r)\left(dx^{0}\right)^2 - \frac{1}{{\cal B}(r)}\delta_{ij}dx^{i} dx^{j}, 
\end{equation} 
where $r=\vert \bm x\vert = \sqrt{\delta_{ij}x^ix^j}$. The potentials ${\cal A}(r)$ and ${\cal B}(r)$ are supposed to satisfy the boundary conditions
\begin{equation} \label{asf}
\lim_{r \rightarrow \infty}{\cal A}(r)=1, \qquad \lim_{r \rightarrow \infty}{\cal B}(r)=1,
\end{equation}
which ensure that the metric is asymptotically flat when $r\rightarrow \infty$. Moreover, we assume that there exists a value $r_h >0$ of the radial coordinate such that the metric is regular for any $r>r_h$. By analogy with the Schwarzschild spacetime, we suppose that $r_h \sim m$. The spatial region corresponding to $r>r_h$ is denoted by ${\cal D}_h$.

Let $\bm x_{\scriptscriptstyle A}$ and $\bm x_{\scriptscriptstyle B}$ be two points located in ${\cal D}_h$. Generically, we consider a photon emitted at $\bm x_{\scriptscriptstyle A}$ at a time $t_{\scriptscriptstyle A}=x^0_{\scriptscriptstyle A}/c$ and received at $\bm x_{\scriptscriptstyle B}$ at a time $t_{\scriptscriptstyle B}=x^0_{\scriptscriptstyle B}/c$. By convention, the instant of reception will be considered as fixed, whereas the instant of emission will depend on the trajectory of the photon. Owing to the static nature of the metric, any mention of $t_{\scriptscriptstyle B}$ will be systematically omitted and the null geodesic path followed by the photon will be denoted by $\Gamma(\bm x_{\scriptscriptstyle A}, \bm x_{\scriptscriptstyle B})$ or simply $\Gamma$ in the absence of ambiguity. A thorough characterization of the direction of propagation of the light at $\bm x_{\scriptscriptstyle A}$ and $\bm x_{\scriptscriptstyle B}$ may be carried out as follows.

Let $x^{\alpha}=x_{\scriptscriptstyle \Gamma}^{\alpha}(\zeta)$ be a system of parametric equations describing $\Gamma(\bm x_{\scriptscriptstyle A}, \bm x_{\scriptscriptstyle B})$, $\zeta$ being an arbitrarily chosen parameter. It is easily seen that the direction of propagation at any point $x$ of $\Gamma(\bm x_{\scriptscriptstyle A}, \bm x_{\scriptscriptstyle B})$ is fully determined by the {\it direction triple} defined as 
\begin{equation} \label{tl}
\widehat{\underline{\bm l}}=\left(l_1 /l_0, l_2 /l_0, l_3 /l_0\right)=\left(l_i /l_0\right),  
\end{equation}
where the quantities $l_0$ and $l_i$ are the covariant components of the vector tangent to the ray at $x$, that is the quantities given by $l_{\alpha}=g_{\alpha\beta}(x_{\scriptscriptstyle \Gamma}^{\rho}(\zeta))dx_{\scriptscriptstyle \Gamma}^{\beta}(\zeta)/d\zeta$. It is worth noting that the direction triple $\widehat{\underline{\bm l}}$ is collinear to the Euclidean 3-vector tangent to the photon trajectory. Indeed a relation as follows
\begin{equation} \label{tlt}
\widehat{\underline{\bm l}}=-\frac{1}{{\cal A}(r) {\cal B}(r)}\frac{d\bm x_{\scriptscriptstyle \Gamma}}{dx^0}
\end{equation}
directly results from (\ref{ds2}) and (\ref{tl}). It is clear that $\widehat{\underline{\bm l}}$ does not depend on the choice of the parameter along $\Gamma(\bm x_{\scriptscriptstyle A}, \bm x_{\scriptscriptstyle B})$.

The expressions of $\widehat{\underline{\bm l}}$ at points $\bm x_{\scriptscriptstyle A}$ and $\bm x_{\scriptscriptstyle B}$ will be denoted by $\widehat{\underline{\bm l}}_{\,e}(\bm x_{\scriptscriptstyle A},\bm x_{\scriptscriptstyle B};\Gamma)$ and $\widehat{\underline{\bm l}}_{\,r}(\bm x_{\scriptscriptstyle A},\bm x_{\scriptscriptstyle B};\Gamma)$, respectively. These triples may be determined, at least in principle, by integrating the null geodesic equations. However, the direction triples at $\bm x_{\scriptscriptstyle A}$ and $\bm x_{\scriptscriptstyle B}$ can also be straightforwardly derived from the relations \cite{Leponcin:2004}
\numparts
\begin{equation} \label{wlA0}
\widehat{\underline{\bm l}}_{\,e}(\bm x_{\scriptscriptstyle A},\bm x_{\scriptscriptstyle B};\Gamma)=\left(\frac{l_i}{l_0}\right)_{\! x_{A}}=\left(c \frac{\partial {\cal T}_{\scriptscriptstyle \Gamma}(\bm x_{\scriptscriptstyle A}, \bm x_{\scriptscriptstyle B})}{\partial x^{i}_{\scriptscriptstyle A}}\right)
\end{equation}
and
\begin{equation}   \label{wlB0}
\widehat{\underline{\bm l}}_{\,r}(\bm x_{\scriptscriptstyle A},\bm x_{\scriptscriptstyle B};\Gamma)=\left(\frac{l_i}{l_0}\right)_{\! x_{B}}=-\left(c \frac{\partial {\cal T}_{\scriptscriptstyle \Gamma}(\bm x_{\scriptscriptstyle A}, \bm x_{\scriptscriptstyle B})}{\partial x^{i}_{\scriptscriptstyle B}}\right),
\end{equation}
\endnumparts
where ${\cal T}_{\scriptscriptstyle \Gamma} (\bm x_{\scriptscriptstyle A}, \bm x_{\scriptscriptstyle B})$ is the so-called time transfer function associated to $\Gamma$, that is the expression yielding the travel time of the photon along $\Gamma$ as a function of the spatial positions of the emitter and the receiver:
\begin{equation} \label{4}
t_{\scriptscriptstyle B} - t_{\scriptscriptstyle A} = {\cal T}_{\scriptscriptstyle \Gamma} (\bm x_{\scriptscriptstyle A}, \bm x_{\scriptscriptstyle B}).
\end{equation}  

Since the metric is static and spherically symmetric,  ${\cal T}_{\scriptscriptstyle \Gamma} (\bm x_{\scriptscriptstyle A}, \bm x_{\scriptscriptstyle B})$ may be considered as a function of $r_{\scriptscriptstyle A}=\vert\bm x_{\scriptscriptstyle A}\vert$, $r_{\scriptscriptstyle B}=\vert\bm x_{\scriptscriptstyle B}\vert$ and the angle between $\bm x_{\scriptscriptstyle A}$ and $\bm x_{\scriptscriptstyle B}$. So defining the unit vectors $\bm n_{\scriptscriptstyle A}$ and $\bm n_{\scriptscriptstyle B}$ by
\begin{equation} \label{n}
\bm n_{\scriptscriptstyle A} = \frac{\bm x_{\scriptscriptstyle A}}{r_{\scriptscriptstyle A}}, \qquad \bm n_{\scriptscriptstyle B} = \frac{\bm x_{\scriptscriptstyle B}}{r_{\scriptscriptstyle B}}, 
\end{equation}
and then setting
\begin{equation} \label{mu}
\mu = \bm n_{\scriptscriptstyle A}.\bm n_{\scriptscriptstyle B},
\end{equation}
one can put 
\[
{\cal T}_{\scriptscriptstyle \Gamma} (\bm x_{\scriptscriptstyle A},\bm x_{\scriptscriptstyle B})={\cal T}_{\scriptscriptstyle \Gamma} (r_{\scriptscriptstyle A}, r_{\scriptscriptstyle B}, \mu). 
\]
As a consequence, equations (\ref{wlA0}) and (\ref{wlB0}) may be written in the form
\numparts
\begin{eqnarray} 
\widehat{\underline{\bm l}}_{\,e}(\bm x_{\scriptscriptstyle A},\bm x_{\scriptscriptstyle B};\Gamma)=c\frac{\partial  {\cal T}_{\scriptscriptstyle \Gamma}}{\partial r_{\scriptscriptstyle A}}\bm n_{\scriptscriptstyle A}+c\frac{\vert \bm n_{\scriptscriptstyle A}\times \bm n_{\scriptscriptstyle B}\vert}{r_{\scriptscriptstyle A}} \frac{\partial  {\cal T}_{\scriptscriptstyle \Gamma}}{\partial \mu} \, \bm Q_{\scriptscriptstyle AB} \times \bm n_{\scriptscriptstyle A}, \label{wlA} \\
\widehat{\underline{\bm l}}_{\,r}(\bm x_{\scriptscriptstyle A},\bm x_{\scriptscriptstyle B};\Gamma)=-c\frac{\partial  {\cal T}_{\scriptscriptstyle \Gamma}}{\partial r_{\scriptscriptstyle B}}\bm n_{\scriptscriptstyle B}+c\frac{\vert \bm n_{\scriptscriptstyle A}\times \bm n_{\scriptscriptstyle B}\vert}{r_{\scriptscriptstyle B}} \frac{\partial  {\cal T}_{\scriptscriptstyle \Gamma}}{\partial \mu} \, \bm Q_{\scriptscriptstyle AB} \times \bm n_{\scriptscriptstyle B}, \label{wlB} 
\end{eqnarray}
\endnumparts
where $\bm Q_{\scriptscriptstyle AB}$ is the Euclidean unit vector defined by
\begin{equation} \label{kAB}
\bm Q_{\scriptscriptstyle AB}=\frac{\bm n_{\scriptscriptstyle A}\times \bm n_{\scriptscriptstyle B}}{\vert\bm n_{\scriptscriptstyle A}\times \bm n_{\scriptscriptstyle B}\vert}.
\end{equation}

It is well known that the angular momentum vector defined as
\begin{equation} \label{L}
\bm L = - \bm x \times \underline{\widehat{\bm l}}
\end{equation}
is conserved along any geodesic of a  static, spherically symmetric spacetime (see appendix A). Let us denote by $\bm L_{\scriptscriptstyle \Gamma}(\bm x_{\scriptscriptstyle A},\bm x_{\scriptscriptstyle B})$ the conserved angular momentum corresponding to the null geodesic path $\Gamma(\bm x_{\scriptscriptstyle A},\bm x_{\scriptscriptstyle B})$. Replacing $\underline{\widehat{\bm l}}$ by $\widehat{\underline{\bm l}}_{\,e}(\bm x_{\scriptscriptstyle A},\bm x_{\scriptscriptstyle B};\Gamma)$ and taking (\ref{wlA}) into account, it is easily seen that $\bm L_{\scriptscriptstyle \Gamma}(\bm x_{\scriptscriptstyle A},\bm x_{\scriptscriptstyle B})$ may be written as 
\begin{equation} \nonumber
\bm L_{\scriptscriptstyle \Gamma}(\bm x_{\scriptscriptstyle A},\bm x_{\scriptscriptstyle B})=b_{\scriptscriptstyle \Gamma}(\bm x_{\scriptscriptstyle A},\bm x_{\scriptscriptstyle B}) \, \bm Q_{\scriptscriptstyle AB},
\end{equation} 
where $b_{\scriptscriptstyle \Gamma}(\bm x_{\scriptscriptstyle A},\bm x_{\scriptscriptstyle B})$ is defined by
\begin{equation} \label{imp3}
b_{\scriptscriptstyle \Gamma}(\bm x_{\scriptscriptstyle A},\bm x_{\scriptscriptstyle B})=-c\frac{\partial {\cal T}_{\scriptscriptstyle \Gamma}}{\partial \mu}\vert\bm n_{\scriptscriptstyle A}\times \bm n_{\scriptscriptstyle B}\vert.
\end{equation}
Finally, using equation (\ref{imp3}), the system of equations (\ref{wlA}) and (\ref{wlB}) reads
\numparts
\begin{eqnarray}
\widehat{\underline{\bm l}}_{\,e}(\bm x_{\scriptscriptstyle A},\bm x_{\scriptscriptstyle B};\Gamma)=c\frac{\partial {\cal T}_{\scriptscriptstyle \Gamma}}{\partial r_{\scriptscriptstyle A}}\bm n_{\scriptscriptstyle A}-\frac{b_{\scriptscriptstyle \Gamma}}{r_{\scriptscriptstyle A}} \bm Q_{\scriptscriptstyle AB} \times \bm n_{\scriptscriptstyle A},  \label{wlA2} \\
\widehat{\underline{\bm l}}_{\,r}(\bm x_{\scriptscriptstyle A},\bm x_{\scriptscriptstyle B};\Gamma)=-c\frac{\partial {\cal T}_{\scriptscriptstyle \Gamma}}{\partial r_{\scriptscriptstyle B}}\bm n_{\scriptscriptstyle B}-\frac{b_{\scriptscriptstyle \Gamma}}{r_{\scriptscriptstyle B}} \bm Q_{\scriptscriptstyle AB} \times \bm n_{\scriptscriptstyle B}.  \label{wlB2}
\end{eqnarray}
\endnumparts

The main interest of introducing systematically the angular momentum vector lies in the fact that the constant of the motion $b_{\scriptscriptstyle \Gamma}$ {\it is an intrinsic quantity}, the magnitude of which may be regarded as the {\em impact parameter} of the ray when conditions (\ref{asf}) are met \cite{Chandrasekhar:1983}. Indeed, for a null geodesic admitting an asymptote, the magnitude of $\bm L$ is such that
$\vert\bm L\vert = lim_{\vert\bm x\vert \rightarrow \infty}\left\vert\bm x\times d\bm x/cdt\right\vert$
since $\underline{\widehat{\bm l}} \longrightarrow -(d\bm x/cdt)_{\infty}$ when $\vert\bm x\vert \longrightarrow \infty$. So $\vert b \vert=\vert-\bm x\times \underline{\widehat{\bm l}}\,\vert $ is the Euclidean distance between the asymptote to the ray and the line parallel to this asymptote passing through the origin of coordinates $O$ as measured by an inertial observer at rest at infinity. 

Equations (\ref{wlA2}) and (\ref{wlB2}) appear to have a quite simple form. However, in order to discuss the problems raised by the bending of light, it will be more convenient to express $\widehat{\underline{\bm l}}_{\,e}$ and $\widehat{\underline{\bm l}}_{\,r}$ as linear combinations of the unit vectors $\bm N_{\scriptscriptstyle AB}$ and $\bm P_{\scriptscriptstyle AB}$ defined by
\begin{equation} \label{NAB}
\bm N_{\scriptscriptstyle AB} = \frac{\bm x_{\scriptscriptstyle B}-\bm x_{\scriptscriptstyle A}}{\vert\bm x_{\scriptscriptstyle B}-\bm x_{\scriptscriptstyle A}\vert}=\frac{r_{\scriptscriptstyle B}}{\vert\bm x_{\scriptscriptstyle B}-\bm x_{\scriptscriptstyle A}\vert} \bm n_{\scriptscriptstyle B}-\frac{r_{\scriptscriptstyle A}}{\vert\bm x_{\scriptscriptstyle B}-\bm x_{\scriptscriptstyle A}\vert}\bm n_{\scriptscriptstyle A}
\end{equation}
and
\begin{equation} \label{PAB}
\bm P_{\scriptscriptstyle AB} = \bm N_{\scriptscriptstyle AB}\times\bm Q_{\scriptscriptstyle AB},
\end{equation}
respectively. Then the triples $\widehat{\underline{\bm l}}_{\,e}$ and $\widehat{\underline{\bm l}}_{\,r}$ may be written as
\numparts
\begin{eqnarray} 
\widehat{\underline{\bm l}}_{\,e}(\bm x_{\scriptscriptstyle A},\bm x_{\scriptscriptstyle B};\Gamma)=&\Bigg\lbrack c\frac{\partial {\cal T}_{\scriptscriptstyle \Gamma}}{\partial r_{\scriptscriptstyle A}}(\bm N_{\scriptscriptstyle AB}.\bm n_{\scriptscriptstyle A})-\frac{b_{\scriptscriptstyle \Gamma}}{r_{\scriptscriptstyle A}}\vert \bm N_{\scriptscriptstyle AB}\times\bm n_{\scriptscriptstyle A}\vert\Bigg\rbrack \bm N_{\scriptscriptstyle AB}  \nonumber \\
&+\Bigg\lbrack c\frac{\partial {\cal T}_{\scriptscriptstyle \Gamma}}{\partial r_{\scriptscriptstyle A}}\vert \bm N_{\scriptscriptstyle AB}\times\bm n_{\scriptscriptstyle A}\vert+\frac{b_{\scriptscriptstyle \Gamma}}{r_{\scriptscriptstyle A}}(\bm N_{\scriptscriptstyle AB}.\bm n_{\scriptscriptstyle A})\Bigg\rbrack \bm P_{\scriptscriptstyle AB}, \label{wlA3} \\
\widehat{\underline{\bm l}}_{\,r}(\bm x_{\scriptscriptstyle A},\bm x_{\scriptscriptstyle B};\Gamma)=&-\Bigg\lbrack c\frac{\partial {\cal T}_{\scriptscriptstyle \Gamma}}{\partial r_{\scriptscriptstyle B}}(\bm N_{\scriptscriptstyle AB}.\bm n_{\scriptscriptstyle B})+\frac{b_{\scriptscriptstyle \Gamma}}{r_{\scriptscriptstyle B}}\vert \bm N_{\scriptscriptstyle AB}\times\bm n_{\scriptscriptstyle B}\vert\Bigg\rbrack \bm N_{\scriptscriptstyle AB} \nonumber \\
&-\Bigg\lbrack c\frac{\partial {\cal T}_{\scriptscriptstyle \Gamma}}{\partial r_{\scriptscriptstyle B}}\vert \bm N_{\scriptscriptstyle AB}\times\bm n_{\scriptscriptstyle B}\vert-\frac{b_{\scriptscriptstyle \Gamma}}{r_{\scriptscriptstyle B}}(\bm N_{\scriptscriptstyle AB}.\bm n_{\scriptscriptstyle B})\Bigg\rbrack \bm P_{\scriptscriptstyle AB}.  \label{wlB3}
\end{eqnarray}
\endnumparts

Equations (\ref{imp3}), (\ref{wlA3}) and (\ref{wlB3}) show that determining the direction triples $\widehat{\underline{\bm l}}_{\,e}$ and $\widehat{\underline{\bm l}}_{\,r}$ comes down to knowing the three quantities $c\partial {\cal T}_{\scriptscriptstyle \Gamma}/\partial \mu$, $c\partial {\cal T}_{\scriptscriptstyle \Gamma}/\partial r_{\scriptscriptstyle A}$ and $c\partial {\cal T}_{\scriptscriptstyle \Gamma}/\partial r_{\scriptscriptstyle B}$ associated to the null geodesic path covered by the photon. However, the problem is far from being solved, since two given points $\bm x_{\scriptscriptstyle A}$ and $\bm x_{\scriptscriptstyle B}$ can generally be linked by an infinite number of null geodesic paths. Furthermore, there exists in general no procedure enabling to determine the time transfer function corresponding to each null geodesic path joining $\bm x_{\scriptscriptstyle A}$ and $\bm x_{\scriptscriptstyle B}$\footnote{This difficulty is facing us even in the Schwarzschild spacetime, in spite of the fact that the exact analytical solution to the differential equations of the null geodesics is known in terms of the Weierstrass elliptic function $\wp$ for a long time (see, e.g., \cite{deJans:1922}).}. As far as we know, these problems can be successfully tackled only for the restricted class constituted by the null geodesic paths that we call below the quasi-Minkowskian light rays. This question will be developed at length in section \ref{SppNap}, after having supplemented the above general considerations by  a remarkably simple expression for the angular distance between two light rays passing through a given point as measured by an observer at rest at this point.

\section{Angle between two light rays observed at the same point} \label{Sangle}

The only quantity having a practical physical meaning in astrometry is the angle between two light rays, say $\Gamma$ and $\Gamma'$, simultaneously arriving at a given point $\bm x_{\scriptscriptstyle B}$ as measured by an observer passing through this point. The aim of the present section is to get a coordinate-independent expression of this angle in any spacetime endowed with a metric given by (\ref{ds2}). For the sake of simplicity, we restrict our attention to the case where the observer is staying at rest at point $\bm x_{\scriptscriptstyle B}$. Such a {\it static observer} will be denoted by ${\cal S}_{\scriptscriptstyle B}$\footnote{The unit 4-velocity vector of ${\cal S}_{\scriptscriptstyle B}$ is collinear to the timelike Killing vector at $\bm x_{\scriptscriptstyle B}$. So the property to be a static observer is an intrinsic one.}. The angle between the rays as measured by another observer passing through $\bm x_{\scriptscriptstyle B}$ will be inferred using, e.g., the formulae given in \cite{Teyssandier:2006} or \cite{Crosta:2010}. 

It is well known that the direction of propagation of a light ray $\Gamma$ as seen by an observer passing through $\bm x_{\scriptscriptstyle B}$ is defined by the projection of a 4-vector $l_{\scriptscriptstyle B}$ tangent to $\Gamma$ at point $\bm x_{\scriptscriptstyle B}$ on the local rest space of this observer. Since the projection of a null vector is always a non-zero spacelike vector, the direction of propagation along $\Gamma$ relative to ${\mathcal S}_{\scriptscriptstyle B}$ is completely defined by the unit 4-vector (see, e.g.,  \cite{Soffel:1989,deFelice:2006,deFelice:2010})
\begin{equation} \label{NU}
n_{\scriptscriptstyle B}=\frac{l_{\scriptscriptstyle B}}{U_{\scriptscriptstyle B}\cdot l_{\scriptscriptstyle B}}-U_{\scriptscriptstyle B},
\end{equation} 
where $U_{\scriptscriptstyle B}$ is the unit 4-velocity vector of ${\mathcal S}_{\scriptscriptstyle B}$, and $U_{\scriptscriptstyle B}\cdot l_{\scriptscriptstyle B}$ denotes the scalar product of $U_{\scriptscriptstyle B}$ and $l_{\scriptscriptstyle B}$ (see section \ref{Snot}). Of course, the direction of $\Gamma'$ arriving at $\bm x_{\scriptscriptstyle B}$ following a tangent vector $l_{\scriptscriptstyle B}'$ will be defined by the 4-vector $N_{\scriptscriptstyle B}'$ obtained by substituting $l_{\scriptscriptstyle B}'$ for $l_{\scriptscriptstyle B}$ into (\ref{NU}). For the sake of brevity, the angular separation between $\Gamma$ and $\Gamma'$ as measured by the static observer ${\mathcal S}_{\scriptscriptstyle B}$ will be henceforth designated by $\delta_{\scriptscriptstyle B}(\Gamma, \Gamma')$ without explicitly mentioning the observer. The value of the angle $\delta_{\scriptscriptstyle B}(\Gamma, \Gamma')$ is fixed without ambiguity by the relation 
\begin{equation} \label{angs1}
\cos \delta_{\scriptscriptstyle B}(\Gamma, \Gamma')= -n_{\scriptscriptstyle B}\cdot n_{\scriptscriptstyle B}'
\end{equation}
supplemented by the condition  
\begin{equation} \label{rgang}
0\leq\delta_{\scriptscriptstyle B}(\Gamma, \Gamma')\leq\pi.
\end{equation}

The notation used for the angles as measured by a given observer must be carefully distinguished from the notation of the Euclidean angles between the direction triples defined in section \ref{Snot}. Surprisingly, however, there exists a very simple connection between the covariantly well-defined angles and the formal Euclidean angles. We have indeed a proposition as follows.

\begin{proposition} \label{Pang}
Let $\Gamma$ and $\Gamma'$ be two light rays arriving at $\bm x_{\scriptscriptstyle B}$ at the same instant. When the metric is written in isotropic coordinates adapted to the symmetries of spacetime, the angle $\delta_{\scriptscriptstyle B}(\Gamma, \Gamma')$ between these rays as measured by a static observer staying at $\bm x_{\scriptscriptstyle B}$ is equal to the Euclidean angle between the triples $\underline{\widehat{\bm l}}_{\,r}$ and  $\widehat{\underline{\bm l}}'_{\,r}$ defining at $\bm x_{\scriptscriptstyle B}$ the directions of propagation along $\Gamma$ and $\Gamma'$, respectively:
\begin{equation} \label{angs2}
\delta_{\scriptscriptstyle B}(\Gamma, \Gamma')=\delta_{\scriptscriptstyle {\cal E}}(\underline{\widehat{\bm l}}_{\,r}, \widehat{\underline{\bm l}}'_{\,r})=
\arccos\left(\frac{\underline{\widehat{\bm l}}_{\,r} .\, \widehat{\underline{\bm l}}'_{\,r}}{\vert\underline{\widehat{\bm l}}_{\,r}\vert \vert\widehat{\underline{\bm l}}'_{\,r}\vert}\right).
\end{equation}
\end{proposition}

{\bf Proof of proposition \ref{Pang}}  $ $ Since the contravariant components of $U_{\scriptscriptstyle B}$ are $U_{\scriptscriptstyle B}^0=1/\sqrt{{\cal A}(r_{\scriptscriptstyle B})}$ and $U_{\scriptscriptstyle B}^i=0$, equation (\ref{NU}) implies that the covariant components of $N_{\scriptscriptstyle B}$ and $N_{\scriptscriptstyle B}'$ are given by $(N_{\scriptscriptstyle B})_0=0, (N_{\scriptscriptstyle B})_i=\sqrt{{\cal A}(r_{\scriptscriptstyle B})}\,(\widehat{l}_{\scriptscriptstyle B})_i$ and $(N_{\scriptscriptstyle B}')_0=0, (N_{\scriptscriptstyle B}')_i=\sqrt{{\cal A}(r_{\scriptscriptstyle B})}\,(\widehat{l}'_{\scriptscriptstyle B})_i$, respectively. Consequently, equation (\ref{angs1}) may be written as
\begin{equation} \label{angs3}
\cos \delta_{\scriptscriptstyle B}(\Gamma, \Gamma')={\cal A}(r_{\scriptscriptstyle B}){\cal B}(r_{\scriptscriptstyle B})\underline{\widehat{\bm l}}_{\,r} .\, \widehat{\underline{\bm l}}'_{\,r}.
\end{equation}
It may be straightforwardly deduced from the property of $l_{\alpha} = g_{\alpha\beta} l^{\beta}$ to be a covariant null vector that 
\begin{equation} \label{nol1}
\vert\underline{\widehat{\bm l}}_{\,r}\vert=\frac{1}{\sqrt{{\cal A}(r_{\scriptscriptstyle B}){\cal B}(r_{\scriptscriptstyle B})}}
\end{equation}
for any null path $\Gamma$. Equation (\ref{angs3}) is therefore equivalent to equation (\ref{angs2}) when condition (\ref{rgang}) is taken into account. QED.

The angular separation $\delta_{\scriptscriptstyle B}(\Gamma, \Gamma')$ is an intrinsic quantity defined as an angle effectively measurable by an observer who is himself defined in a way which does not depend on the choice of coordinates. So the equality of $\delta_{\scriptscriptstyle B}(\Gamma, \Gamma')$ with a purely formal Euclidean angle $\delta_{\scriptscriptstyle {\cal E}}(\underline{\widehat{\bm l}}_{\,r}, \widehat{\underline{\bm l}}'_{\,r})$ was not obvious. In fact this equality would be broken in any coordinate system differing from the isotropic one. The numerical coincidence yielded by proposition \ref{Pang} therefore confirms the relevance of our choice of coordinates.

\section{Post-post-Minkowskian approximation} \label{SppNap}

From now on we assume that in the exterior region ${\cal D}_h$ the functions ${\cal A}(r)$ and ${\cal B}(r)^{-1}$ admit analytic expansions in powers of $m/r$ as follow
\numparts
\begin{eqnarray}
{\cal A}(r)= 1 - \frac{2m}{r} + 2\beta \frac{m^2}{r^2} + \cdots , \label{A1} \\
\frac{1}{{\cal B}(r)}=1 + 2 \gamma \frac{m}{r} + \frac{3}{2}\epsilon \frac{m^2}{r^2} + \cdots , \label{C1}
\end{eqnarray}
\endnumparts
where $\beta$ and $\gamma$ are the usual post-Newtonian parameters, and $\epsilon$ is a post-post-Newtonian parameter ($\beta = \gamma = \epsilon = 1$ in general relativity). Since $m$ is proportional to the gravitational constant $G$, equations (\ref{A1}) and (\ref{C1}) may be considered as yielding a post-post-Minkowskian expansion of the metric. In agreement with this assumption, we may suppose that given an arbitrary reception point $\bm x_{\scriptscriptstyle B}$ in ${\cal D}_h$ there exists a large region of ${\cal D}_h$ such that any point $\bm x_{\scriptscriptstyle A}$ in this region is linked to $\bm x_{\scriptscriptstyle B}$ by one and only one null geodesic path which is a perturbation in powers of $G$ of a Minkowskian null geodesic. We may admit that such a configuration correspond to almost all the situations effectively encountered in astrometry. To be more precise we shall adopt the following definition. 

\begin{definition} \label{qMink}
A null geodesic path joining two points $\bm x_{\scriptscriptstyle A}$ and $\bm x_{\scriptscriptstyle B}$ in the domain of analyticity ${\cal D}_h$ is said to be a quasi-Minkowskian light ray if this path lies in ${\cal D}_h$ and can be described by parametric equations which take the form
\begin{eqnarray} 
x^{0}=x^0_{\scriptscriptstyle B}-\lambda \vert \bm x_{\scriptscriptstyle B}-\bm x_{\scriptscriptstyle A}\vert +\sum_{n=1}^{\infty}G^n X^0_{(n)}(\bm x_{\scriptscriptstyle A}, \bm x_{\scriptscriptstyle B}, \lambda) , \label{qM0} \\
\bm x=\bm x_{\scriptscriptstyle B}-\lambda (\bm x_{\scriptscriptstyle B}-\bm x_{\scriptscriptstyle A})+\sum_{n=1}^{\infty}G^n \bm X_{(n)}(\bm x_{\scriptscriptstyle A}, \bm x_{\scriptscriptstyle B}, \lambda),  \label{qMi}
\end{eqnarray}
where $\lambda$ is the affine parameter varying on the range $0\leq \lambda \leq 1$, with the boundary conditions
\begin{equation} \label{bcdt}
 X^0_{(n)}(\bm x_{\scriptscriptstyle A}, \bm x_{\scriptscriptstyle B}, 0)=0
\end{equation}
and
\begin{equation} \label{bcd}
\bm X_{(n)}(\bm x_{\scriptscriptstyle A}, \bm x_{\scriptscriptstyle B}, 0)=\bm X_{(n)}(\bm x_{\scriptscriptstyle A}, \bm x_{\scriptscriptstyle B}, 1)=0
\end{equation}
being met whatever $n\geq 1$.
\end{definition}

It is shown in \cite{Teyssandier:2008} that the time transfer function corresponding to a quasi-Minkowskian light ray is yielded as a series in powers of $G$ by an iterative procedure involving only integrations along the segment defined by parametric equations $\bm x = \bm x_{\scriptscriptstyle B}-\lambda (\bm x_{\scriptscriptstyle B}-\bm x_{\scriptscriptstyle A})$, with $0\leq \lambda \leq 1$. This procedure implies the uniqueness of the solution. The path between $\bm x_{\scriptscriptstyle A}$ and $\bm x_{\scriptscriptstyle B}$ corresponding to definition \ref{qMink} will be henceforth denoted by $\Gamma_{s}(\bm x_{\scriptscriptstyle A}, \bm x_{\scriptscriptstyle B})$ or simply by $\Gamma_{s}$\footnote{The index $s$ is adopted in order to recall that a quasi-Minkowskian null geodesic path may be considered as an almost {\it straight} line.}. For the sake of brevity, we shall drop any mention of path $\Gamma_{s}$ in the notations previously introduced for the time transfer function and the impact parameter. Thus ${\cal T}(\bm x_{\scriptscriptstyle A}, \bm x_{\scriptscriptstyle B})$ and $b$ shall henceforth stand for ${\cal T}_{\scriptscriptstyle \Gamma_{s}} (\bm x_{\scriptscriptstyle A}, \bm x_{\scriptscriptstyle B})$ and $b_{\scriptscriptstyle \Gamma_{s}}(\bm x_{\scriptscriptstyle A}, \bm x_{\scriptscriptstyle B})$, respectively. Moreover, we shall systematically omit any mention of $\Gamma_{s}$ in the arguments of the direction triples.

Since each term in the perturbation expansion of ${\cal T}(\bm x_{\scriptscriptstyle A}, \bm x_{\scriptscriptstyle B})$ is an integral involving the values of the metric tensor on the segment joining $\bm x_{\scriptscriptstyle A}$ and $\bm x_{\scriptscriptstyle B}$, we shall assume that the condition 
\begin{equation} \label{val1}
\vert\bm x_{\scriptscriptstyle B}-\lambda (\bm x_{\scriptscriptstyle B}-\bm x_{\scriptscriptstyle A})\vert > r_h\quad \mbox{when}\quad 0\leq \lambda \leq 1
\end{equation}
is fulfilled in order to guarantee the validity of the expressions we use. It is worthy of note that condition (\ref{val1}) is differently ensured according to the signs of $\bm N_{\scriptscriptstyle AB}.\bm n_{\scriptscriptstyle A}$ and $\bm N_{\scriptscriptstyle AB}.\bm n_{\scriptscriptstyle B}$:

{\em a}) If $\bm N_{\scriptscriptstyle AB}.\bm n_{\scriptscriptstyle A} \geq 0$, (\ref{val1}) is met if and only if $r_{\scriptscriptstyle A}>r_h$;

{\em b}) If $\bm N_{\scriptscriptstyle AB}.\bm n_{\scriptscriptstyle A} < 0$ and $\bm N_{\scriptscriptstyle AB}.\bm n_{\scriptscriptstyle B} \leq 0$, (\ref{val1}) is met if and only if $r_{\scriptscriptstyle B}>r_h$;

{\em c}) If $\bm N_{\scriptscriptstyle AB}.\bm n_{\scriptscriptstyle A} <0$ and $\bm N_{\scriptscriptstyle AB}.\bm n_{\scriptscriptstyle B} > 0$, (\ref{val1}) holds provided that
\begin{equation} \label{rch}
r_c>r_h,
\end{equation}
where $r_c$ is the so-called {\it zeroth-order distance of closest approach}, that is the Euclidean distance of the straight line passing through $\bm x_{\scriptscriptstyle A}$ and $\bm x_{\scriptscriptstyle B}$ from the origin $O$: 
\begin{equation} \label{rc1}
r_c=\frac{r_{\scriptscriptstyle A} r_{\scriptscriptstyle B}\vert \bm n_{\scriptscriptstyle A}\times\bm n_{\scriptscriptstyle B}\vert}{\vert\bm x_{\scriptscriptstyle B}-\bm x_{\scriptscriptstyle A}\vert}.
\end{equation}

We shall not examine here if the results established below might be considered as valid under a less constraining condition than (\ref{val1}).

It is shown in \cite{Leponcin:2004} and \cite{Teyssandier:2008} that the time transfer function corresponding to $\Gamma_s(\bm x_{\scriptscriptstyle A}, \bm x_{\scriptscriptstyle B})$ is given by\footnote{The term of order $G$ in (\ref{Tss}) is the well-known Shapiro time delay (see, e.g., \cite{Will:1993} and references therein).} 
\begin{eqnarray}
\fl\mathcal{T}(\bm x_{\scriptscriptstyle A},\bm x_{\scriptscriptstyle B})=\frac{\vert\bm x_{\scriptscriptstyle B}-\bm x_{\scriptscriptstyle A}\vert}{c}+\frac{(1+\gamma)m}{c}
\ln\left(\frac{r_{\scriptscriptstyle A}+r_{\scriptscriptstyle B}+\vert\bm x_{\scriptscriptstyle B}-\bm x_{\scriptscriptstyle A}\vert}{r_{\scriptscriptstyle A}+r_{\scriptscriptstyle B}-\vert\bm x_{\scriptscriptstyle B}-\bm x_{\scriptscriptstyle A}\vert}\right)\nonumber \\
+\frac{m^{2}}{r_{\scriptscriptstyle A} r_{\scriptscriptstyle B}}\frac{\vert\bm x_{\scriptscriptstyle B}-\bm x_{\scriptscriptstyle A}\vert}{c}\left[\kappa
\frac{\arccos (\bm n_{\scriptscriptstyle A}.\bm n_{\scriptscriptstyle B})}
{\vert\bm n_{\scriptscriptstyle A}\times \bm n_{\scriptscriptstyle B}\vert}
-\frac{(1+\gamma)^2}{1+\bm n_{\scriptscriptstyle A} . \bm n_{\scriptscriptstyle B}}\right]+ O(G^3),  \label{Tss}
\end{eqnarray}
where $\kappa$ is the dimensionless parameter defined by
\begin{equation} \label{kappa}
\kappa=\mbox{$\frac{1}{4}$}[8(1+\gamma)-4\beta+3\epsilon].
\end{equation}

It may be noted that $\kappa = 15/4$ in general relativity. 

The impact parameter of $\Gamma_s(\bm x_{\scriptscriptstyle A}, \bm x_{\scriptscriptstyle B})$ is easy to determine. Substituting for ${\cal T} (\bm x_{\scriptscriptstyle A}, \bm x_{\scriptscriptstyle B})$ from (\ref{Tss}) into (\ref{imp3}), noting that 
\[
\vert\bm x_{\scriptscriptstyle B}-\bm x_{\scriptscriptstyle A}\vert=(r_{\scriptscriptstyle A}^2+r_{\scriptscriptstyle B}^2-2r_{\scriptscriptstyle A}r_{\scriptscriptstyle B}\mu)^{1/2},
\]
and then using the relations
\begin{equation} \label{rel1}
\frac{r_{\scriptscriptstyle A}\vert\bm n_{\scriptscriptstyle A}\times\bm n_{\scriptscriptstyle B}\vert}{\vert\bm x_{\scriptscriptstyle B}-\bm x_{\scriptscriptstyle A}\vert}=\vert\bm N_{\scriptscriptstyle AB}\times\bm n_{\scriptscriptstyle B}\vert, \qquad \frac{r_{\scriptscriptstyle B}\vert\bm n_{\scriptscriptstyle A}\times\bm n_{\scriptscriptstyle B}\vert}{\vert\bm x_{\scriptscriptstyle B}-\bm x_{\scriptscriptstyle A}\vert}=\vert\bm N_{\scriptscriptstyle AB}\times\bm n_{\scriptscriptstyle A}\vert
\end{equation}
and
\begin{equation} \label{id1}
\bm n_{\scriptscriptstyle A}.\bm n_{\scriptscriptstyle B} - \vert\bm N_{\scriptscriptstyle AB}\times\bm n_{\scriptscriptstyle A}\vert \vert\bm N_{\scriptscriptstyle AB}\times\bm n_{\scriptscriptstyle B}\vert= (\bm N_{\scriptscriptstyle AB}.\bm n_{\scriptscriptstyle A}) (\bm N_{\scriptscriptstyle AB}.\bm n_{\scriptscriptstyle B}),
\end{equation}
a straightforward calculation leads to the following proposition.

\begin{proposition} \label{Prb1}
Let $\bm x_{\scriptscriptstyle A}$ and $\bm x_{\scriptscriptstyle B}$ be two points such that condition (\ref{val1}) is fulfilled. The quantity $b$ associated to a quasi-Minkowskian light ray $\Gamma_s(\bm x_{\scriptscriptstyle A},\bm x_{\scriptscriptstyle B})$ by equation (\ref{imp3}) is given by 
\begin{eqnarray}
\fl b= r_c\Bigg\lbrace 1+\frac{(1+\gamma)m}{r_c} \frac{\vert\bm N_{\scriptscriptstyle AB}\times\bm n_{\scriptscriptstyle A}\vert+\vert\bm N_{\scriptscriptstyle AB}\times\bm n_{\scriptscriptstyle B}\vert}{1 + \bm n_{\scriptscriptstyle A}.
\bm n_{\scriptscriptstyle B}} \nonumber \\ 
+\frac{m^2}{r_c^2}\Bigg\lbrace \kappa\left[1-\frac{\arccos(\bm n_{\scriptscriptstyle A} . \bm n_{\scriptscriptstyle B})}{\vert \bm n_{\scriptscriptstyle A}\times\bm n_{\scriptscriptstyle B}\vert}(\bm N_{\scriptscriptstyle AB}.\bm n_{\scriptscriptstyle A})(\bm N_{\scriptscriptstyle AB}.\bm n_{\scriptscriptstyle B})\right] \nonumber \\
-(1+\gamma)^2 \frac{1 - (\bm N_{\scriptscriptstyle AB}.\bm n_{\scriptscriptstyle A}) (\bm N_{\scriptscriptstyle AB}.\bm n_{\scriptscriptstyle B})}{1 + \bm n_{\scriptscriptstyle A}.\bm n_{\scriptscriptstyle B}}\Bigg\rbrace\Bigg\rbrace+O\left(\frac{m^3}{r_c^2}\right), \label{imp1}
\end{eqnarray}
where $r_c$ is the zeroth-order distance of closest approach defined by equation (\ref{rc1}). 
\end{proposition}

Since the right-hand side of (\ref{imp1}) is a positive quantity when $r_c \gg m$, we shall henceforth identify $b$ with the impact parameter of the quasi-Minkowskian ray $\Gamma_s(\bm x_{\scriptscriptstyle A},\bm x_{\scriptscriptstyle B})$.

We are now in a position to determine the triples $\underline{\widehat{\bm l}}_{\,e}$ and $\underline{\widehat{\bm l}}_{\,r}$ corresponding to $\Gamma_s(\bm x_{\scriptscriptstyle A},\bm x_{\scriptscriptstyle B})$ by substituting for ${\cal T}(\bm x_{\scriptscriptstyle A}, \bm x_{\scriptscriptstyle B})$ from (\ref{Tss}) into (\ref{wlA3}) and (\ref{wlB3}), respectively. Using once again (\ref{rel1}), taking (\ref{imp1}) into account and using the relation
\begin{equation} \label{id2}
\vert \bm n_{\scriptscriptstyle A}\times\bm n_{\scriptscriptstyle B}\vert +\vert\bm N_{\scriptscriptstyle AB}\times\bm n_{\scriptscriptstyle B}\vert \bm N_{\scriptscriptstyle AB}.\bm n_{\scriptscriptstyle A}-\vert\bm N_{\scriptscriptstyle AB}\times\bm n_{\scriptscriptstyle A}\vert \bm N_{\scriptscriptstyle AB}.\bm n_{\scriptscriptstyle B}=0,
\end{equation} 
a straightforward calculation leads to the following proposition.

\begin{proposition} \label{DirAB0}
Under the assumption of proposition \ref{Prb1}, the triples $\underline{\widehat{\bm l}}_{\,e}$ and $\underline{\widehat{\bm l}}_{\,r}$ corresponding to a quasi-Minkowskian light ray $\Gamma_s(\bm x_{\scriptscriptstyle A},\bm x_{\scriptscriptstyle B})$ admit the expansions in powers of $m/r_c$ given by 
\numparts
\begin{eqnarray}
\fl\underline{\widehat{\bm l}}_{\,e}(\bm x_{\scriptscriptstyle A},\bm x_{\scriptscriptstyle B})=-\bm N_{\scriptscriptstyle AB}\nonumber \\
-\frac{m \vert\bm N_{\scriptscriptstyle AB}\times\bm n_{\scriptscriptstyle A}\vert}{r_{c}}\Bigg\lbrace 1+\gamma+\frac{m \vert\bm N_{\scriptscriptstyle AB}\times\bm n_{\scriptscriptstyle A}\vert}{r_{c}}\Bigg\lbrack\kappa-\frac{(1+\gamma)^2}{1+\bm n_{\scriptscriptstyle A}.\bm n_{\scriptscriptstyle B}}\Bigg\rbrack\Bigg\rbrace\bm N_{\scriptscriptstyle AB}\nonumber \\
-\frac{m \vert\bm N_{\scriptscriptstyle AB}\times\bm n_{\scriptscriptstyle A}\vert}{r_{c}}\,\Bigg\lbrace(1+\gamma)\frac{\vert\bm n_{\scriptscriptstyle A}\times\bm n_{\scriptscriptstyle B}\vert}{1+\bm n_{\scriptscriptstyle A}.\bm n_{\scriptscriptstyle B}}+\frac{m}{r_{c}}\Bigg\lbrace \kappa\Bigg\lbrack \frac{\arccos(\bm n_{\scriptscriptstyle A}.\bm n_{\scriptscriptstyle B})}{\vert\bm n_{\scriptscriptstyle A}\times\bm n_{\scriptscriptstyle B}\vert}\bm N_{\scriptscriptstyle AB}.\bm n_{\scriptscriptstyle B}\nonumber \\
-\bm N_{\scriptscriptstyle AB}.\bm n_{\scriptscriptstyle A}\Bigg\rbrack -(1+\gamma)^2 \,\frac{\bm N_{\scriptscriptstyle AB}.\bm n_{\scriptscriptstyle B}-\bm N_{\scriptscriptstyle AB}.\bm n_{\scriptscriptstyle A}}{1+\bm n_{\scriptscriptstyle A}.\bm n_{\scriptscriptstyle B}}\Bigg\rbrace\Bigg\rbrace\bm P_{\scriptscriptstyle AB}+O\left(\frac{m^3}{r_{c}^3}\right)\label{dirA2}
\end{eqnarray}
and
\begin{eqnarray}
\fl\underline{\widehat{\bm l}}_{\,r}(\bm x_{\scriptscriptstyle A},\bm x_{\scriptscriptstyle B})=-\bm N_{\scriptscriptstyle AB}\nonumber \\
-\frac{m \vert\bm N_{\scriptscriptstyle AB}\times\bm n_{\scriptscriptstyle B}\vert}{r_{c}}\Bigg\lbrace 1+\gamma+\frac{m \vert\bm N_{\scriptscriptstyle AB}\times\bm n_{\scriptscriptstyle B}\vert}{r_{c}}\Bigg\lbrack\kappa-\frac{(1+\gamma)^2}{1+\bm n_{\scriptscriptstyle A}.\bm n_{\scriptscriptstyle B}}\Bigg\rbrack\Bigg\rbrace\bm N_{\scriptscriptstyle AB}\nonumber\\
+\frac{m \vert\bm N_{\scriptscriptstyle AB}\times\bm n_{\scriptscriptstyle B}\vert}{r_{c}}\,\Bigg\lbrace(1+\gamma)\frac{\vert\bm n_{\scriptscriptstyle A}\times\bm n_{\scriptscriptstyle B}\vert}{1+\bm n_{\scriptscriptstyle A}.\bm n_{\scriptscriptstyle B}}-\frac{m}{r_{c}}\Bigg\lbrace \kappa\Bigg\lbrack \frac{\arccos(\bm n_{\scriptscriptstyle A}.\bm n_{\scriptscriptstyle B})}{\vert\bm n_{\scriptscriptstyle A}\times\bm n_{\scriptscriptstyle B}\vert}\bm N_{\scriptscriptstyle AB}.\bm n_{\scriptscriptstyle A}\nonumber \\
-\bm N_{\scriptscriptstyle AB}.\bm n_{\scriptscriptstyle B}\Bigg\rbrack +(1+\gamma)^2 \,\frac{\bm N_{\scriptscriptstyle AB}.\bm n_{\scriptscriptstyle B}-\bm N_{\scriptscriptstyle AB}.\bm n_{\scriptscriptstyle A}}{1+\bm n_{\scriptscriptstyle A}.\bm n_{\scriptscriptstyle B}}\Bigg\rbrace\Bigg\rbrace\bm P_{\scriptscriptstyle AB}+O\left(\frac{m^3}{r_{c}^3}\right),\label{dirB2}
\end{eqnarray}
\endnumparts
respectively.
\end{proposition}

Owing to the intrinsic nature of the impact parameter, it is relevant to form the expansions of $\underline{\widehat{\bm l}}_{\,e}$ and $\underline{\widehat{\bm l}}_{\,r}$ in powers of $m/b$. We just have to replace $1/r_c$ into (\ref{dirA2}) and (\ref{dirB2}) by its expansion inferred from (\ref{imp1}), namely 
\begin{equation} \label{rcimp}
\frac{1}{r_c}=\frac{1}{b}\Bigg\lbrack 1+\frac{(1+\gamma)m}{b} \frac{\vert\bm N_{\scriptscriptstyle AB}\times\bm n_{\scriptscriptstyle A}\vert+\vert\bm N_{\scriptscriptstyle AB}\times\bm n_{\scriptscriptstyle B}\vert}{1 + \bm n_{\scriptscriptstyle A}.
\bm n_{\scriptscriptstyle B}}\Bigg\rbrack + O\left(\frac{m^2}{b^3}\right).
\end{equation}
Using the relation
\begin{equation} \label{id3}
\left[\vert\bm N_{\scriptscriptstyle AB}\times\bm n_{\scriptscriptstyle A}\vert+\vert\bm N_{\scriptscriptstyle AB}\times\bm n_{\scriptscriptstyle B}\vert\right]\frac{\vert \bm n_{\scriptscriptstyle A}\times\bm n_{\scriptscriptstyle B}\vert}{1+\bm n_{\scriptscriptstyle A}.\bm n_{\scriptscriptstyle B}}+\bm N_{\scriptscriptstyle AB}.\bm n_{\scriptscriptstyle A}-\bm N_{\scriptscriptstyle AB}.\bm n_{\scriptscriptstyle B}=0
\end{equation}
we obtain the fundamental expressions below.

\begin{proposition} \label{DirAB}
The triples $\underline{\widehat{\bm l}}_{\,e}$ and $\underline{\widehat{\bm l}}_{\,r}$ yielded by proposition \ref{DirAB0} may be rewritten in the form
\numparts
\begin{eqnarray}
\fl\underline{\widehat{\bm l}}_{\,e}(\bm x_{\scriptscriptstyle A},\bm x_{\scriptscriptstyle B})=-\bm N_{\scriptscriptstyle AB}-\frac{m\vert\bm N_{\scriptscriptstyle AB}\times\bm n_{\scriptscriptstyle A}\vert}{b}\Bigg\lbrace 1+\gamma
+\frac{m}{b}\Bigg\lbrack \kappa\vert\bm N_{\scriptscriptstyle AB}\times\bm n_{\scriptscriptstyle A}\vert \nonumber \\
+ (1+\gamma)^2 \frac{\vert\bm N_{\scriptscriptstyle AB}\times\bm n_{\scriptscriptstyle B}\vert}{1+\bm n_{\scriptscriptstyle A}.\bm n_{\scriptscriptstyle B}}\Bigg\rbrack\Bigg\rbrace\bm N_{\scriptscriptstyle AB} \nonumber \\
-\frac{m\vert\bm N_{\scriptscriptstyle AB}\times\bm n_{\scriptscriptstyle A}\vert}{b}\Bigg\lbrace(1+\gamma)\frac{\vert\bm n_{\scriptscriptstyle A}\times\bm n_{\scriptscriptstyle B}\vert}{1+\bm n_{\scriptscriptstyle A}.\bm n_{\scriptscriptstyle B}}\nonumber \\
+\frac{\kappa m}{b}\Bigg\lbrack\frac{\arccos(\bm n_{\scriptscriptstyle A}.\bm n_{\scriptscriptstyle B})}{\vert\bm n_{\scriptscriptstyle A}\times\bm n_{\scriptscriptstyle B}\vert}\bm N_{\scriptscriptstyle AB}.\bm n_{\scriptscriptstyle B}-\bm N_{\scriptscriptstyle AB}.\bm n_{\scriptscriptstyle A}\Bigg\rbrack\Bigg\rbrace\bm P_{\scriptscriptstyle AB}+O\left(\frac{m^3}{b^3}\right), \label{dirA2c}\\
\fl\underline{\widehat{\bm l}}_{\,r}(\bm x_{\scriptscriptstyle A},\bm x_{\scriptscriptstyle B})=-\bm N_{\scriptscriptstyle AB}-\frac{m\vert\bm N_{\scriptscriptstyle AB}\times\bm n_{\scriptscriptstyle B}\vert}{b}\Bigg\lbrace 1+\gamma+\frac{m}{b}\Bigg\lbrack \kappa\vert\bm N_{\scriptscriptstyle AB}\times\bm n_{\scriptscriptstyle B}\vert \nonumber \\
+ (1+\gamma)^2 \frac{\vert\bm N_{\scriptscriptstyle AB}\times\bm n_{\scriptscriptstyle A}\vert}{1+\bm n_{\scriptscriptstyle A}.\bm n_{\scriptscriptstyle B}}\Bigg\rbrack\Bigg\rbrace\bm N_{\scriptscriptstyle AB} \nonumber \\
+\frac{m\vert\bm N_{\scriptscriptstyle AB}\times\bm n_{\scriptscriptstyle B}\vert}{b}\,\Bigg\lbrace(1+\gamma)\frac{\vert\bm n_{\scriptscriptstyle A}\times\bm n_{\scriptscriptstyle B}\vert}{1+\bm n_{\scriptscriptstyle A}.\bm n_{\scriptscriptstyle B}}\nonumber \\
-\frac{\kappa m}{b}\Bigg\lbrack\frac{\arccos(\bm n_{\scriptscriptstyle A}.\bm n_{\scriptscriptstyle B})}{\vert\bm n_{\scriptscriptstyle A}\times\bm n_{\scriptscriptstyle B}\vert}\bm N_{\scriptscriptstyle AB}.\bm n_{\scriptscriptstyle A}-\bm N_{\scriptscriptstyle AB}.\bm n_{\scriptscriptstyle B}\Bigg\rbrack\Bigg\rbrace\bm P_{\scriptscriptstyle AB}+O\left(\frac{m^3}{b^3}\right), \label{dirB2c}
\end{eqnarray}
\endnumparts
where $b$ is the impact parameter of the ray given by equation (\ref{imp1}).
\end{proposition}

Equations (\ref{dirA2c}) and (\ref{dirB2c}) have clearly a more simple form than equations (\ref{dirA2}) and (\ref{dirB2}), respectively.

It may be noted that the Euclidean norm $\vert \underline{\widehat{\bm l}}_{\,r} \vert$ is now given by 
\begin{equation} \label{nol2}
\vert\underline{\widehat{\bm l}}_{\,r}\vert = 
1 + \frac{(1+\gamma)m}{r_{\scriptscriptstyle B}} + \frac{1}{2}\left[2\kappa - (1+\gamma)^2\right]\frac{m^2}{r_{\scriptscriptstyle B}^2} + O\left(\frac{m^3}{r_{\scriptscriptstyle B}^3}\right),
\end{equation}
which implies that the factor $1/\vert\underline{\widehat{\bm l}}_{\,r}\vert.\vert\widehat{\underline{\bm l}}'_{\,r}\vert$ 
involved in (\ref{angs2}) reads
\begin{equation} \label{expAB}
\frac{1}{\vert\underline{\widehat{\bm l}}_{\,r}\vert.\vert\widehat{\underline{\bm l}}'_{\,r}\vert}=1 - \frac{2(1+\gamma)m}{r_{\scriptscriptstyle B}} - 2\left[\kappa - 2(1+\gamma)^2\right]\frac{m^2}{r_{\scriptscriptstyle B}^2} + O\left(\frac{m^3}{r_{\scriptscriptstyle B}^3}\right).
\end{equation}

\section{Case of a ray emitted at infinity} \label{Sinfty}

The notion of quasi-Minkowskian light ray can be extended as follows when the ray is emitted at infinity. 

\begin{definition} \label{qMinkb}
Let $\bm N_e$ be a unit 3-vector and $\bm x_{\scriptscriptstyle B}$ a point in ${\cal D}_h$. A null geodesic path is said to be a quasi-Minkowskian light ray emitted at infinity in the direction $\bm N_e$ and received at point $\bm x_{\scriptscriptstyle B}$ if this path entirely lies in ${\cal D}_h$ and admits parametric equations having the form
\begin{eqnarray}
x^0=x^0_{\scriptscriptstyle B}+\sigma-\sigma_{\scriptscriptstyle B}+\sum_{n=1}^{\infty} G^n X_{(n)}^0(\bm N_e, \bm x_{\scriptscriptstyle B}, \sigma),  \label{qM0b} \\
\bm x=\bm x_{\scriptscriptstyle B}+(\sigma -\sigma_{\scriptscriptstyle B})\bm N_e +\sum_{n=1}^{\infty} G^n \bm X_{(n)}(\bm N_e, \bm x_{\scriptscriptstyle B}, \sigma),  \label{qMib}
\end{eqnarray} 
where $\sigma$ is an affine parameter varying in the range $-\infty < \sigma \leq \sigma_{\scriptscriptstyle B}$, the boundary conditions  
\begin{equation} \label{bcdbt}
X_{(n)}^0(\bm N_e, \bm x_{\scriptscriptstyle B}, \sigma_{\scriptscriptstyle B})=0
\end{equation}
and
\begin{equation} \label{bcdbx}
\bm X_{(n)}(\bm N_e, \bm x_{\scriptscriptstyle B}, \sigma_{\scriptscriptstyle B})=0
\end{equation}
being fulfilled.
\end{definition}

Such a ray will be denoted by $\Gamma_{s}(\bm N_e , \bm x_{\scriptscriptstyle B})$. Its impact parameter\footnote{It is natural to assume that $\big\vert\sum_{n=1}^{\infty} G^n \bm X_{(n)}(\bm N_e, \bm x_{\scriptscriptstyle B}, \sigma).
[\bm n_{\scriptscriptstyle B}-(\bm N_e.\bm n_{\scriptscriptstyle B})\bm N_e]\big\vert$ generically admits a bounded limit when $\sigma\rightarrow -\infty$. This assumption means that the ray is asymptotic to a straight line parallel to $\bm N_e$.} $b$ is given by equation (\ref{imp1}) in which $\bm x_{\scriptscriptstyle A}$ may be any arbitrarily chosen point on $\Gamma_{s}(\bm N_e , \bm x_{\scriptscriptstyle B})$. Indeed, it is easily seen that introducing the affine parameter $\lambda=(\sigma_{\scriptscriptstyle B}-\sigma)/(\sigma_{\scriptscriptstyle B}-\sigma_{\scriptscriptstyle A})$ transforms the part of $\Gamma_{s}(\bm N_e , \bm x_{\scriptscriptstyle B})$ joining $\bm x_{\scriptscriptstyle A}$ and $\bm x_{\scriptscriptstyle B}$ into a null geodesic path satisfying all the criteria set in definition \ref{qMink}. Consequently, $b$ may be obtained as a function of $\bm N_e$ and $\bm x_{\scriptscriptstyle B}$ by taking the limit of the right-hand side of (\ref{imp1})  when $r_{\scriptscriptstyle A}\rightarrow \infty$, $\bm n_{\scriptscriptstyle A}\rightarrow -\bm N_e$ and $\bm N_{\scriptscriptstyle AB}\rightarrow \bm N_e$. It will be useful to note that $\arccos(\bm n_{\scriptscriptstyle A}\cdot\bm n_{\scriptscriptstyle B}) \rightarrow \pi - \arccos(\bm N_e \cdot \bm n_{\scriptscriptstyle B})$ when $\bm x_{\scriptscriptstyle A}$ is moved away along the ray towards infinity. It may be pointed out, however, that a condition like (\ref{val1}) has to be fulfilled by $\bm x_{\scriptscriptstyle A}$ in order to guarantee the validity of (\ref{imp1}). Unfortunately, it seems impossible to know if such a condition is effectively met by any point of $\Gamma_{s}(\bm N_e , \bm x_{\scriptscriptstyle B})$. So we shall henceforth content ourselves with assuming the validity of the asymptotic limit of condition (\ref{val1}), namely  
\begin{equation} \label{stDh}
\vert(\sigma-\sigma_{\scriptscriptstyle B})\bm N_e + \bm x_{\scriptscriptstyle B}\vert>r_h \quad\mbox{when}\quad -\infty<\sigma\leq\sigma_{\scriptscriptstyle B}.
\end{equation}
Condition (\ref{stDh}) means that the straight segment parallel to $\bm N_e$ and ending at point $\bm x_{\scriptscriptstyle B}$ is entirely inside the domain ${\cal D}_h$. A proposition as follows can be formulated.

\begin{proposition} \label{Prb2}
Let $\bm N_e$ be a unit vector and $\bm x_{\scriptscriptstyle B}$ a point in ${\cal D}_h$ chosen so that condition (\ref{stDh}) is met. The impact parameter $b$ of a quasi-Minkowkian light ray $\Gamma_{s}(\bm N_e , \bm x_{\scriptscriptstyle B})$ is given by
\begin{eqnarray}
\fl b=r_c\Bigg\lbrace 1+\frac{(1+\gamma)m}{r_c} \frac{\vert\bm N_{e}\times\bm n_{\scriptscriptstyle B}\vert}{1 - \bm N_{e}.
\bm n_{\scriptscriptstyle B}} \nonumber \\ 
+\frac{m^2}{r_c^2}\Bigg\lbrace\kappa\left[1+\frac{\pi-\arccos(\bm N_{e} . \bm n_{\scriptscriptstyle B})}{\vert \bm N_{e}\times\bm n_{\scriptscriptstyle B}\vert}\bm N_{e}.\bm n_{\scriptscriptstyle B}\right]-(1+\gamma)^2 \frac{1 + \bm N_{e}.\bm n_{\scriptscriptstyle B}}{1 - \bm N_{e}.\bm n_{\scriptscriptstyle B}}\Bigg\rbrace\Bigg\rbrace\nonumber \\
+O\left(\frac{m^3}{r_c^2}\right), \label{imp4}
\end{eqnarray}
where $r_c$ is the zeroth-order distance of closest approach defined as
\begin{equation} \label{rc}
r_{c}=r_{\scriptscriptstyle B}\vert \bm N_e\times\bm n_{\scriptscriptstyle B}\vert.
\end{equation} 
\end{proposition}

A similar reasoning applied to (\ref{dirB2}) yields the expression of the triple $\underline{\widehat{\bm l}}_{\,r}(\bm N_e , \bm x_{\scriptscriptstyle B})$ characterizing the direction of $\Gamma_{s}(\bm N_e , \bm x_{\scriptscriptstyle B})$ at $\bm x_{\scriptscriptstyle B}$ as a function of $\bm N_e$ and $\bm x_{\scriptscriptstyle B}$.
\begin{proposition} \label{dirBrc}
Under the assumptions of proposition \ref{Prb2}, the triple yielding the direction of propagation of a quasi-Minkowskian light ray $\Gamma_{s}(\bm N_e , \bm x_{\scriptscriptstyle B})$ at point $\bm x_{\scriptscriptstyle B}$ is given by  
\begin{eqnarray}
\fl\underline{\widehat{\bm l}}_{\,r}(\bm N_e , \bm x_{\scriptscriptstyle B})=-\bm N_{e}\nonumber \\
-\frac{m}{r_{c}}\Bigg\lbrace (1+\gamma)\vert\bm N_{e}\times\bm n_{\scriptscriptstyle B}\vert+\frac{m}{r_{c}}\bigg\lbrack\kappa\vert\bm N_{e}\times\bm n_{\scriptscriptstyle B}\vert^2-(1+\gamma)^2(1+\bm N_{e}.\bm n_{\scriptscriptstyle B})\bigg\rbrack\Bigg\rbrace\bm N_{e}\nonumber\\
+\frac{m}{r_{c}}\,\Bigg\lbrace(1+\gamma)(1+\bm N_{e}.\bm n_{\scriptscriptstyle B})+\frac{m}{r_{c}}\Bigg\lbrace \kappa\bigg\lbrack \pi - \arccos(\bm N_{e}.\bm n_{\scriptscriptstyle B})\nonumber \\
+\vert\bm N_{e}\times\bm n_{\scriptscriptstyle B}\vert\bm N_{e}.\bm n_{\scriptscriptstyle B}\bigg\rbrack
-(1+\gamma)^2 \,\frac{(1+\bm N_{e}.\bm n_{\scriptscriptstyle B})^2}{\vert\bm N_{e}\times\bm n_{\scriptscriptstyle B}\vert}\Bigg\rbrace\Bigg\rbrace\bm P_{e}+O\left(\frac{m^3}{r_{c}^3}\right),\label{dirB3a}
\end{eqnarray}
where $\bm P_{e}$ is the unit 3-vector orthogonal to $\bm N_e$ defined as
\begin{equation} \label{PAB2}
\bm P_{e}=\frac{\bm N_{e}\times\bm n_{\scriptscriptstyle B}}{\vert\bm N_{e}\times\bm n_{\scriptscriptstyle B}\vert} \times \bm N_{e}.
\end{equation}
\end{proposition}

Of course, the unit vector $\bm N_e$ and the impact parameter $b$ may be taken as the boundary values which determine a quasi-Minkowskian light ray emitted at infinity. Then the notation $\Gamma_{s}(\bm N_e , b)$ instead of $\Gamma_{s}(\bm N_e , \bm x_{\scriptscriptstyle B})$ will be used. The following proposition is straightforwardly deduced from (\ref{dirB2c}). 

\begin{proposition} \label{dirB}
Let $\Gamma_{s}(\bm N_e , b)$ be a quasi-Minkowskian light ray emitted at infinity in the direction $\bm N_e$ with an impact parameter $b$. At any point $\bm x_{\scriptscriptstyle B}$ of $\Gamma_{s}(\bm N_e , b)$ fulfilling condition (\ref{stDh}), the direction triple is given by  
\begin{eqnarray}
\fl \underline{\widehat{\bm l}}_{\,r}(\bm N_e , \bm x_{\scriptscriptstyle B})=-\bm N_{e}-\frac{m\vert\bm N_{e}\times\bm n_{\scriptscriptstyle B}\vert}{b}\Bigg\lbrack 1+\gamma+\frac{\kappa m\vert\bm N_{e}\times\bm n_{\scriptscriptstyle B}\vert}{b}\Bigg\rbrack\bm N_{e} \nonumber \\
+\frac{m}{b}\Bigg\lbrace (1+\gamma)(1+\bm N_{e}.\bm n_{\scriptscriptstyle B})+\frac{\kappa m}{b}\bigg\lbrack\pi-\arccos(\bm N_{e}.\bm n_{\scriptscriptstyle B})\nonumber \\
+\vert\bm N_{e}\times\bm n_{\scriptscriptstyle B}\vert \bm N_{e}.\bm n_{\scriptscriptstyle B}\bigg\rbrack\Bigg\rbrace
\bm P_{e}+O\left(\frac{m^3}{b^3}\right).\label{dirB3}
\end{eqnarray}
\end{proposition}

We note that using an expansion in powers of $m/b$ instead of $m/r_c$ cancels the `enhanced' term $-(1+\gamma)^2 (m/r_c)^2(1+\bm N_{e}.\bm n_{\scriptscriptstyle B})^2/\vert\bm N_{e}\times\bm n_{\scriptscriptstyle B}\vert$ appearing in the right-hand side of (\ref{dirB3a}). We shall discuss this question in the next section.

\section{Bending of light rays coming from infinity} \label{Sbend}
 
In order to define the gravitational deflection of a light ray relative to a static observer ${\mathcal S}_{\scriptscriptstyle B}$ staying at a given point $\bm x_{\scriptscriptstyle B}$, let us consider first two quasi-Minkowskian light rays $\Gamma_s$ and $\Gamma'_s$ arriving at the same instant at $\bm x_{\scriptscriptstyle B}$. We suppose that $\Gamma_s$ and $\Gamma'_s$ are emitted at infinity in initial directions $\bm N_e$ and $\bm N'_e$, respectively. Furthermore, we assume that a condition like (\ref{stDh}) is met by each of these rays. If gravity was lacking, the angle between $\Gamma_s$ and $\Gamma'_s$ as measured by ${\mathcal S}_{\scriptscriptstyle B}$ would coincide with the Euclidean angle $\delta_{\scriptscriptstyle {\cal E}}(\bm N_e,\bm N'_e) $ formed by $\bm N_e$ and $\bm N'_e$.  The value of this angle may be considered as an intrinsic quantity since $\bm N_e$ and $\bm N'_e$ could be determined (at least in principle) by measurements performed by static observers located at spatial infinity. As a consequence, the gravitational contribution to the angular separation between $\Gamma_s$ and $\Gamma'_s$ as measured by ${\mathcal S}_{\scriptscriptstyle B}$ may be characterized in an intrinsic way by a definition as follows.

\begin{definition} \label{Dangdis}
Suppose that a static observer at point $\bm x_{\scriptscriptstyle B}$ is receiving two quasi-Minkowskian light rays $\Gamma_s$ and $\Gamma'_s$ emitted at infinity in the directions defined by the unit vectors $\bm N_e$ and $\bm N'_e$, respectively. The gravitational contribution to the angular separation between $\Gamma_s$ and $\Gamma'_s$ as measured by this observer is by definition the angular quantity 
\begin{equation} \label{DgNN}
\delta_{\scriptscriptstyle B}^{(gr)} (\Gamma_s, \Gamma'_s)=\delta_{\scriptscriptstyle B} (\Gamma_s, \Gamma'_s) - \delta_{\scriptscriptstyle {\cal E}}(\bm N_e , \bm N'_e).
\end{equation}

\end{definition}

Taking (\ref{angs3}) into account, it is immediately seen that $\delta_{\scriptscriptstyle B}^{(gr)} (\Gamma_s, \Gamma'_s)$ may be determined by the relation
\begin{equation} \label{gdef}
\fl\delta_{\scriptscriptstyle B}^{(gr)} (\Gamma_s, \Gamma'_s)=\arccos\left[{\cal A}(r_{\scriptscriptstyle B}){\cal B}(r_{\scriptscriptstyle B})\,\underline{\widehat{\bm l}}_{\,r}(\bm N_e,\bm x_{\scriptscriptstyle B}) .\widehat{\underline{\bm l}}'_{\,r}(\bm N'_e,\bm x_{\scriptscriptstyle B})\right]-\arccos(\bm N_e .\bm N'_e).
\end{equation}

Definition \ref{Dangdis} and equation (\ref{DgNN}) enable to obtain a coordinate-independent characterization of the gravitational deflection of a quasi-Minkowskian light ray coming from infinity. There exists one and only one null radial geodesic emitted at infinity and arriving at $\bm x_{\scriptscriptstyle B}$ at the instant of reception of $\Gamma_s$. Pointing out the quasi-Minkowskian nature of this ingoing null radial geodesic, we may use definition \ref{Dangdis} in order to propose the new definition below.

\begin{definition} \label{deflN}
Let $\Gamma_s$ be a quasi-Minkowskian light ray emitted at infinity in a given direction $\bm N_e$ and observed at a point $\bm x_{\scriptscriptstyle B}$ fulfilling condition (\ref{stDh}). The gravitational deflection of this ray relative to a static observer at $\bm x_{\scriptscriptstyle B}$ is by definition the angular quantity 
\begin{equation} \label{gde1}
\Delta\chi_{\scriptscriptstyle B}(\Gamma_s)=-\delta_{\scriptscriptstyle B}^{(gr)} (\Gamma_s, \Gamma_{rd}^{-}),
\end{equation}
where $\Gamma_{rd}^{-}$ is the ingoing radial null geodesic path arriving at $\bm x_{\scriptscriptstyle B}$ at the instant of reception of $\Gamma_s$.
\end{definition}  

The sign occurring in the right-hand-side of (\ref{gde1}) is chosen in order to have a positive quantity for the light deflection. The ingoing null radial geodesic $\Gamma_{rd}^{-}$ is emitted at infinity in the direction given by the unit vector $-\bm n_{\scriptscriptstyle B}$. So the direction triple of this geodesic at $\bm x_{\scriptscriptstyle B}$ may be denoted by $\widehat{\underline{\bm l}}_{r}(-\bm n_{\scriptscriptstyle B}, \bm x_{\scriptscriptstyle B})$ according to the notation already used in section \ref{Sinfty}. A closed-form, exact expression of $\widehat{\underline{\bm l}}_{r}(-\bm n_{\scriptscriptstyle B}, \bm x_{\scriptscriptstyle B})$ is easily derived from the geodesic equations. One gets
\begin{equation} \label{dirnB}
\widehat{\underline{\bm l}}_{r}(-\bm n_{\scriptscriptstyle B}, \bm x_{\scriptscriptstyle B})=\frac{\bm n_{\scriptscriptstyle B}}{\sqrt{{\cal A}(r_{\scriptscriptstyle B}){\cal B}(r_{\scriptscriptstyle B})}}.
\end{equation}
Taking (\ref{dirnB}) into account, it is easily seen that (\ref{angs2}) implies $\delta_{\scriptscriptstyle B} (\Gamma_s, \Gamma^{-}_{rd})=\delta_{\scriptscriptstyle {\cal E}}(\widehat{\underline{\bm l}}_{\,r}(\bm N_{e} , \bm x_{\scriptscriptstyle B}), \bm n_{\scriptscriptstyle B})$. Hence an equation as follows, directly inferred from equations (\ref{DgNN}) and (\ref{gde1}):
\begin{equation} \label{gde1a}
\Delta\chi_{\scriptscriptstyle B}(\Gamma_s)=\delta_{\scriptscriptstyle {\cal E}}(-\bm N_e , \bm n_{\scriptscriptstyle B}) - \delta_{\scriptscriptstyle {\cal E}}(\widehat{\underline{\bm l}}_{\,r}(\bm N_{e} , \bm x_{\scriptscriptstyle B}), \bm n_{\scriptscriptstyle B}).
\end{equation}
Since $\widehat{\underline{\bm l}}_{\,r}(\bm N_{e} , \bm x_{\scriptscriptstyle B})$, $\bm n_{\scriptscriptstyle B}$ and $\bm N_{e}$ are coplanar triples, one has 
\begin{equation} \label{deNn}
\delta_{\scriptscriptstyle {\cal E}}(-\bm N_e , \bm n_{\scriptscriptstyle B})=\delta_{\scriptscriptstyle {\cal E}}(-\bm N_e , \widehat{\underline{\bm l}}_{\,r}(\bm N_{e} , \bm x_{\scriptscriptstyle B}))+\delta_{\scriptscriptstyle {\cal E}}(\widehat{\underline{\bm l}}_{\,r}(\bm N_{e} , \bm x_{\scriptscriptstyle B}), \bm n_{\scriptscriptstyle B}).
\end{equation}
Substituting for $\delta_{\scriptscriptstyle {\cal E}}(-\bm N_e , \bm n_{\scriptscriptstyle B})$ from (\ref{deNn}) into (\ref{gde1a}), we obtain the following proposition.
\begin{proposition} \label{Pgde2}
The gravitational deflection of a quasi-Minkowskian light ray $\Gamma_s(\bm N_e , \bm x_{\scriptscriptstyle B})$ relative to a static observer at $\bm x_{\scriptscriptstyle B}$ is equal to the Euclidean angle between  $-\bm N_e$ and the direction triple of the ray at $\bm x_{\scriptscriptstyle B}$:
\begin{equation} \label{gde2}
\Delta\chi_{\scriptscriptstyle B}(\Gamma_s)=\delta_{\scriptscriptstyle {\cal E}}(-\bm N_e , \widehat{\underline{\bm l}}_{\,r}(\bm N_{e} , \bm x_{\scriptscriptstyle B})).
\end{equation}
\end{proposition}

In any practical case encountered in high-accuracy astrometric projects, the gravitational deflection is a small positive angle. As a consequence, (\ref{gde2}) is equivalent to
\begin{equation} \label{gde3}
\Delta\chi_{\scriptscriptstyle B}(\Gamma_s)=\arcsin\left(\sqrt{{\cal A}(r_{\scriptscriptstyle B}){\cal B}(r_{\scriptscriptstyle B})}\,\vert\underline{\widehat{\bm l}}_{\,r}(\bm N_{e} , \bm x_{\scriptscriptstyle B})\times\bm N_{e}\vert\right).
\end{equation}

Equations (\ref{DgNN})-(\ref{gde3}) are exact relations. Within the post-post-Minkowskian approximation, however, the terms of order $G^3$ may be neglected. So the expression of the gravitational deflection reduces to
\begin{equation} \label{chiB1}
\Delta\chi_{\scriptscriptstyle B}(\Gamma_s)=\sqrt{{\cal A}(r_{\scriptscriptstyle B}){\cal B}(r_{\scriptscriptstyle B})}\,\vert\underline{\widehat{\bm l}}_{\,r}(\bm N_{e} , \bm x_{\scriptscriptstyle B})\times\bm N_{e}\vert+O\left(\frac{m^3}{b^3}\right).
\end{equation}

Substituting for $\underline{\widehat{\bm l}}_{\,r}(\bm N_{e} , \bm x_{\scriptscriptstyle B})$ from (\ref{dirB3}) into (\ref{chiB1}), and then replacing $m$ by $GM/c^2$ in order to facilitate precise numerical estimates, we get a proposition as follows.

\begin{proposition} \label{Pdefl}
Let $\Gamma_{s}$ be a quasi-Minkowskian light ray emitted at infinity in a given direction $\bm N_{e}$ and observed at a point $\bm x_{\scriptscriptstyle B}$ fulfilling condition (\ref{stDh}). The gravitational deflection of $\Gamma_{s}$ relative to a static observer at $\bm x_{\scriptscriptstyle B}$ is given by 
\begin{eqnarray} 
\fl \Delta\chi_{\scriptscriptstyle B}(\Gamma_s)=\frac{(1+\gamma)GM}{c^2 b}(1+\cos\phi_{\scriptscriptstyle B})\nonumber \\
+\frac{G^2M^2}{c^4 b^2}\Bigg\lbrack\kappa\!\left(\pi-\phi_{\scriptscriptstyle B}+\frac{1}{2}\sin2\phi_{\scriptscriptstyle B}\right)-(1+\gamma)^2(1+\cos\phi_{\scriptscriptstyle B})\sin\phi_{\scriptscriptstyle B}\Bigg\rbrack \nonumber \\
+O\left(\frac{G^3M^3}{c^6 b^3}\right), \label{chiB2}
\end{eqnarray}
where $b$ is the impact parameter of the ray determined by equation (\ref{imp4}) and $\phi_{\scriptscriptstyle B}$ is the Euclidean angle formed by $\bm N_{e}$ and $\bm n_{\scriptscriptstyle B}$, namely the angle defined by 
\begin{equation} \label{phiB}
\bm N_{e}.\bm n_{\scriptscriptstyle B}=\cos\phi_{\scriptscriptstyle B},\qquad  0\leq\phi_{\scriptscriptstyle B}\leq\pi.
\end{equation}
\end{proposition}

The term of order $G$ in (\ref{chiB2}) is currently used in VLBI astrometry. The contribution of order $G^2$ may amount to about 5.5 microarcseconds ($\mu$as) for a ray coming from infinity and arriving tangentially at the surface of the Sun. So the terms in $G^2$ will be indispensable in a foreseeable future in order to perform highly precise tests of general relativity with laser rays passing near the Sun (LATOR or ASTROD projects, e.g.). 

The right-hand side of equation (\ref{chiB2}) yields a coordinate-independent expression of the deflection of light which involves only two quantities, each of them having a perfectly clear geometrical meaning. Indeed, $b$ is a length which could be measured by a static observer at infinity, and $\phi_{\scriptscriptstyle B}$ is the angle formed by two directions defined in the rest space at infinity. Nevertheless, the intrinsic nature of $\Delta\chi_{\scriptscriptstyle B}(\Gamma_s)$ does not automatically prevent some problems from raising in the discussion of the different contributions. To see that, let us replace $b$ by its expression yielded by equation (\ref{imp4}). Then $\Delta\chi_{\scriptscriptstyle B}(\Gamma_s)$ reads
\begin{eqnarray}
\fl \Delta\chi_{\scriptscriptstyle B}(\Gamma_s)=\frac{(1+\gamma)GM}{c^2 r_c}(1+\cos\phi_{\scriptscriptstyle B})\nonumber \\
+\frac{G^2M^2}{c^4 r_c^2}\Bigg\lbrace\kappa \left(\pi - \phi_{\scriptscriptstyle B}+\frac{1}{2}\sin2\phi_{\scriptscriptstyle B}\right)\nonumber \\
-(1+\gamma)^2 (1+\cos\phi_{\scriptscriptstyle B})\Bigg\lbrack \sin\phi_{\scriptscriptstyle B}+\frac{1+\cos\phi_{\scriptscriptstyle B}}{\sin\phi_{\scriptscriptstyle B}}\Bigg\rbrack\Bigg\rbrace + O\left(\frac{G^3M^3}{c^6 r_c^3}\right), \label{chiB3}
\end{eqnarray}  
where $r_c=r_{\scriptscriptstyle B}\sin\phi_{\scriptscriptstyle B}$. Let us suppose that we apply (\ref{chiB3}) to a ray deflected by a massive body of the Solar System and observed by Gaia, which occupies an orbit about 1.5 million kilometres from the Earth around the Lagrange point L2. Denoting by $r_{\scriptscriptstyle P}$ the value of $r$ at the periapsis of the ray, we have $r_c \approx r_{\scriptscriptstyle P}$ and $\sin\phi_{\scriptscriptstyle B}\approx r_{\scriptscriptstyle P}/r_{\scriptscriptstyle B}$. Then the term in $1/\sin\phi_{\scriptscriptstyle B}$ appearing on the right-hand side of equation (\ref{chiB3}) seemingly generates a post-post-Minkowskian contribution 
\begin{equation} \label{enh}
\Delta\chi^{\scriptscriptstyle (2,enh)}_{\scriptscriptstyle B} \approx- (1+\gamma)^2 (1+\cos \phi_{\scriptscriptstyle B})^2\frac{G^2M^2}{c^4r_{\scriptscriptstyle P}^2} \frac{r_{\scriptscriptstyle B}}{r_{\scriptscriptstyle P}},
\end{equation}
which is equivalent to the so-called `enhanced' post-post-Newtonian term given by equation (94) of \cite{Klioner:2010}. The magnitude of this contribution is totally negligible when the deflecting body is the Sun. Indeed, the fact that in this case $r_{\scriptscriptstyle B}\approx 1$ AU and $r_{\scriptscriptstyle P}> 0.7$ AU for any light ray observed by Gaia implies $\vert\Delta\chi^{\scriptscriptstyle (2,enh)}_{\scriptscriptstyle B}\vert_{Sun} < 6.7 \times 10^{-4} \mu \mbox{as}$. Nevertheless, the right-hand side of (\ref{enh}) may be comparatively large, e.g., for a light ray grazing Jupiter; one has indeed in this case $4 \mbox{AU}\leq r_{\scriptscriptstyle B} \leq 6 \mbox{AU}$ whereas $r_{\scriptscriptstyle P} \approx 7.14  \times 10^4$ km. Then $10.7 \, \mu \mbox{as}\leq \vert\Delta\chi^{\scriptscriptstyle (2,enh)}_{\scriptscriptstyle B}\vert_{Jupiter}\leq 16 \, \mu \mbox{as}$, which means that the value of the effect may be slightly greater than the standard error about 10 $\mu$as expected for the brightest stars during the Gaia mission \cite{deBruijne:2012}. This effect is illusory, however, since $r_c$ differs from $r_{\scriptscriptstyle P}$ by a term of order $G$ which compensates the contribution (\ref{enh}), as it is shown by the absence of any term in $1/\sin\phi_{\scriptscriptstyle B}$ in the right-hand side of equation (\ref{chiB2}). So the `enhanced' term $\Delta\chi^{\scriptscriptstyle (2,enh)}_{\scriptscriptstyle B}$ must be considered as a fictitious contribution due to the fact that neither $r_c$ nor $r_{\scriptscriptstyle P}$ are intrinsically defined parameters. This analysis confirms the conclusions drawn in \cite{Klioner:2010} and \cite{Zschocke:2010}.

\section{Total deflection of light}\label{Stdef}

It is now easy to carry out the calculation of the total deflection $\Delta\chi$ of a quasi-Minkowskian light ray $\Gamma_s$ emitted at infinity with an impact parameter $b\gg r_h$ and returning to infinity after passing near the central body. Since condition (\ref{stDh}) cannot be fulfilled for any point $\bm x_{\scriptscriptstyle B}$ on $\Gamma_s$, we shall restrict our attention on the portion of $\Gamma_s$ between the emission point at infinity and the periapsis of the ray, say $\bm x_{\scriptscriptstyle P}$. Owing to the symmetry, we can put 
\begin{equation} \label{chi}
\Delta\chi = 2 \Delta\chi_{\scriptscriptstyle P}(\Gamma_s),
\end{equation}
where $\Delta\chi_{\scriptscriptstyle P}(\Gamma_s)$ denotes the gravitational deflection of the light ray $\Gamma_s(\bm N_e, \bm x_{\scriptscriptstyle P})$ relative to a static observer at $\bm x_{\scriptscriptstyle P}$, $\bm N_e$ being the emission direction of $\Gamma_s$ at infinity. Since $\underline{\widehat{\bm l}}_{r}(\bm N_e, \bm x_{\scriptscriptstyle P}).\bm n_{\scriptscriptstyle P}=0$, we have 
\begin{equation} \label{angP}
\delta_{\scriptscriptstyle {\cal E}}(\widehat{\underline{\bm l}}_{r}(\bm N_{e} , \bm x_{\scriptscriptstyle P}), \bm n_{\scriptscriptstyle P})=\pi/2.
\end{equation}
Substituting for $\delta_{\scriptscriptstyle {\cal E}}(\widehat{\underline{\bm l}}_{r}(\bm N_{e} , \bm x_{\scriptscriptstyle P}), \bm n_{\scriptscriptstyle P})$ from (\ref{angP}) into (\ref{gde1a}) written at $\bm x_{\scriptscriptstyle P}$, noting that $\delta_{\scriptscriptstyle {\cal E}}(-\bm N_e , \bm n_{\scriptscriptstyle P})=\pi - \phi_{\scriptscriptstyle P}$, where $\phi_{\scriptscriptstyle P}$ is the Euclidean angle between $\bm N_e$ and $\bm n_{\scriptscriptstyle P}$, and then taking (\ref{chi}) into account, we find  
\begin{equation} \label{Dchi1}
\phi_{\scriptscriptstyle P}=\frac{\pi}{2}-\frac{\Delta\chi}{2}.
\end{equation}
Replacing now $\Delta\chi_{\scriptscriptstyle B}(\Gamma_s)$ by $\Delta\chi_{\scriptscriptstyle P}(\Gamma_s)$ and $\phi_{\scriptscriptstyle B}$ by $\phi_{\scriptscriptstyle P}$ in (\ref{chiB2}), and then taking into account (\ref{chi}) and (\ref{Dchi1}), we get 
\begin{eqnarray} 
\fl\Delta\chi =\frac{2(1+\gamma)GM}{c^2 b}\left(1+\sin\frac{\Delta\chi}{2}\right)
+\frac{G^2M^2}{c^4b^2}\Bigg\lbrack\kappa\left(\pi+\Delta\chi+\sin\Delta\chi\right)\nonumber \\
-2(1+\gamma)^2\left(1+\sin\frac{\Delta\chi}{2}\right)\cos\frac{\Delta\chi}{2}\Bigg\rbrack 
+O\left(\frac{G^3\!M^3}{c^6 b^3}\right).\label{chiB4}
\end{eqnarray}
Finally, solving (\ref{chiB4}) for $\Delta\chi$, and then neglecting all terms of order higher than $m^2/b^2$, we are led to the proposition which follows.

\begin{proposition} \label{Ptotdfl}
The total gravitational deflection $\Delta\chi$ of a quasi-Minkowskian light ray starting from infinity with an impact parameter $b\gg r_h$ is given by 
\begin{equation} \label{defl}
\Delta\chi=\frac{2(1+\gamma)GM}{c^2b}+\frac{\kappa\pi G^2M^2}{c^4b^2}+O\left(\frac{G^3M^3}{c^6b^3}\right).
\end{equation}
\end{proposition}

Equation (\ref{defl}) agrees with the results classically yielded by integrating the null geodesic equations (see, e.g., \cite{Klioner:2010} and references therein). 

\section{Conclusion} \label{Conclusion}

In this paper, the direction of light propagation is obtained at the post-post-Minkowskian level for a three-parameter family of static, spherically symmetric spacetimes without integrating the geodesic equations. It must be emphasized, however, that the explicit expansions obtained in sections \ref{SppNap}-\ref{Stdef} only hold for the null geodesic paths that we have called the quasi-Minkowskian light rays. It may be noted that this limitation is also encountered in the previous works treating the problem by integrating the differential equations of null geodesics. 

Our method based on the time transfer function is perfectly adapted to the generic case where the emitter and the receiver are both located at a finite distance from the source. The central result of the paper is constituted by equations (\ref{dirA2}) and (\ref{dirB2}) yielding explicit expressions for the direction triples of a quasi-Minkowskian light ray at its emission and reception points. Supplemented by the formula (\ref{imp1}) yielding the impact parameter $b$ of the ray, these equations straightforwardly lead to equations (\ref{dirA2c}) and (\ref{dirB2c}) giving the expansions of the direction triples in powers of $m/b$. 

The expansions obtained in the generic case are easily extended to a light ray emitted at infinity, the receiver being an observer staying at a finite distance from the centre. An intrinsic characterization of the gravitational deflection angle of such a ray observed by a static observer is formulated in definition \ref{Dangdis}. The well-known expression of the deflection angle currently used in VLBI astrometry is extended to order $G^2$ by equation (\ref{chiB2}). A recent discussion relative to an apparent `enhanced' post-post-Minkowskian term is confirmed. Finally, the well-known formula giving the total light deflection angle up to order $G^2$ is recovered. 

\ack

We are deeply grateful to Prof. B. Linet for critically reading the manuscript and making a lot of useful remarks. 

\appendix

\section{Conservation of the angular momentum} \label{AppA}

In order to prove that the angular momentum $\bm L$ defined by equation (\ref{L}) is conserved along any geodesic path of the metric (\ref{ds2}), it may be assumed without lack of generality that $\zeta$ is an affine parameter. The geodesic paths is then governed by the system of Euler-Lagrange equations
\begin{eqnarray}
& &\frac{dl_{0}}{d\zeta}=0, \label{EL0} \\
& &\frac{dl_{i}}{d\zeta}=\frac{1}{2}\left[{\cal A}'(r)(l^0)^2+\frac{{\cal B}'(r)}{{\cal B}^2(r)}\delta_{kl}l^{k}l^{l}\right]\frac{x^i}{r}, \label{EL1}
\end{eqnarray}
where $l^{\mu}=dx^{\mu}/d\zeta$. Equation (\ref{EL0}) implies
\begin{equation} \label{l0ct}
l_0 \equiv {\cal A}(r)\frac{dx^0}{d\zeta}=E,
\end{equation}
where $E=\mbox{const}$. Excluding now the spacelike geodesics located in an hypersurface $x^0=\mbox{const}$, we can suppose that $E\neq0$. So dividing (\ref{EL1}) by $l_0$ and taking (\ref{l0ct}) into account, it is easily seen that the geodesic path satisfies an equation as follows
\begin{equation} \label{EL2}
\frac{d\underline{\widehat{{\bm l}}}}{d\zeta}=\frac{E}{2}\left[\frac{{\cal A}'(r)}{{\cal A}^2(r)}+{\cal B}'(r)\left(\underline{\widehat{{\bm l}}}\right)^2\right]\frac{\bm x}{r},
\end{equation}
where $\underline{\widehat{{\bm l}}}$ is defined by equation (\ref{tl}). Equations (\ref{tlt}) and (\ref{EL2}) imply the conservation law $d(-\bm x\times\widehat{{\bm l}})/d\zeta=0$.

\end{document}